%% file: ase.tex
\documentclass[conference]{IEEEtran}
\ifCLASSINFOpdf
\else
\fi

\usepackage{multicol,graphicx,amssymb,mathrsfs,amsmath,pifont,amscd,latexsym,amsthm,color,fancyhdr,CJK,url,hyperref,longtable, pifont, subfigure,epstopdf,setspace,multirow,colortbl,tabularx,threeparttable, booktabs, verbatim, bbm,listings, balance,color,listings}
\usepackage[linesnumbered,boxed,algosection]{algorithm2e}

\usepackage[T1]{fontenc}   
\usepackage{txfonts}
\usepackage{mathptmx}
\usepackage{xcolor}

\newcommand{\tabincell}[2]{\begin{tabular}{@{}#1@{}}#2\end{tabular}}

\begin{document}
%
\title{DroidWalker: Generating Reproducible Test Cases via Automatic Exploration of Android Apps}

\author{\IEEEauthorblockN{Ziniu Hu, Yun Ma, Yangyang Huang, Xuanzhe Liu}
\IEEEauthorblockA{Key Laboratory of High Confidence Software Technologies (Peking University), Ministry of Education, Beijing, China \\
Email: \{bull, mayun, hyy\_01, liuxuanzhe\}@pku.edu.cn}
}


%


\maketitle

\begin{abstract}
\input{section/abstract}
\end{abstract}


%
\IEEEpeerreviewmaketitle

\section{Introduction}\label{sec:introduction}
\input{section/introduction}

\section{Related Work}\label{sec:related}
\input{section/relatedwork}

\section{Approach}\label{sec:approach}
\input{section/tool}

\section{Usage Scenario}\label{sec:usage}
\input{section/Usage_scenarios}

\section{Conclusion}\label{sec:conclusion}
\input{section/conclusion}

\bibliographystyle{abbrv}
\bibliography{ase}

\end{document}

%% file: section/abstract.tex
Generating test cases through automatic app exploration is very useful for analyzing and testing Android apps. However, test cases generated by current app-exploration tools are not reproducible, i.e. when the generated test case is re-executed, the app cannot reach the same state as the explored one. As a result, app developers are not able to reproduce the failure or crash reported during the exploration, to conduct regression test after fixing the bug, or to execute the same test in different environments. In this paper, we present DroidWalker, a dynamic-analysis tool to generate reproducible test cases for Android apps. The key design of our tool is a dynamic-adaptive model that can abstract the app state in a proper granularity so every state in the model can be reached afterwards. Given an app under test, DroidWalker first explores the app to build the model. Then developers can select the state in the model to be reproduced. Finally, DroidWalker executes all the generated test cases and the app could reach exactly the same state as the explored one. We apply DroidWalker in three real usage scenarios to demonstrate its practical usage. The video of our tool is at \url{https://youtu.be/ndUD8Gxs800}.

%% file: section/introduction.tex
Nowadays, apps have achieved great success on mobile devices. With increasing number of apps, mobile users spend most of their digital time on apps. Just like all the software, apps should be sufficiently tested before being delivered to end users. Generating test cases via automatically exploring the app under test is a promising method to test the app in the real-world environment. Researchers and practitioners have designed and implemented lots of app exploration tools for Android apps~\cite{Choudhary2016Automated} given the open-source nature and high popularity of Android platform. By triggering events under specific strategies, these tools traverse the app under test for developers to examine app's behavior.

A key requirement of test cases is reproducible, i.e. when the test case is re-executed, the app should reach the same state as the one when the test case is created. The property of reproducibility is important because developers can 1) reproduce the failures and examine the causes for repairing; 2) conduct regression tests after fixing bugs; and 3) examine the consistency of app behaviors in different environments.

However, as pointed out by Choudhary \textit{et al.}~\cite{Choudhary2016Automated}, none of the state-of-the-art automatic app-exploration tools can generate reproducible test cases. The key reason is that these tools do not properly address the dynamics of apps. After exploration, given an app state to be reproduced, if we simply trigger the event trace recorded in the exploration phase, the target app state cannot always be reached. During the re-execution, the app content may be changed so that each event may be triggered on different widgets leading to different app states from those in the exploration phase. For example, when testing a news application, sometimes a news update notification appears on top of the news list, bringing down all the news boxes. Clicking on the previously recorded screen location may not enter the same news page as explored.

To address the issue, in this paper, we present DroidWalker\footnote{The tool and all of our experimental infrastructure and data are publicly available at \url{http://sei.pku.edu.cn/~mayun11/droidwalker/index.html}}, an app exploration tool that can generate reproducible test cases for Android apps. The key design of DroidWalker is a dynamic-adaptive model by which minimal number of events that can tolerate trivial UI changes are generated so that the target states can be reproduced. Given an app under test, DroidWalker first explores the app to build the model. Then developers can select the state to be reproduced and DroidWalker generates test cases that can reach the selected state. Finally, DroidWalker executes all the generated test cases and the app could reach exactly the same state as the explored one.

Apart from generating reproducible test cases, as an app exploration tool, DroidWalker also has the following advantages:
\begin{itemize}
\item DroidWalker achieves higher coverage than existing model-based exploration tools.
\item DroidWalker is able to explore commercial apps and generate test cases.
\item DroidWalker provides interfaces for developers to customize monitoring tasks, being highly configurable.
\end{itemize}

The remainder of this paper is organized as follows. Section~\ref{sec:related} surveys existing tools. Section~\ref{sec:approach} describes the design and implementation of DroidWalker. Section~\ref{sec:usage} presents three usage scenarios to demonstrate DroidWalker. Section~\ref{sec:conclusion} draws a brief conclusion.

%% file: section/relatedwork.tex
Although generating reproducible test cases for Android apps is essential to developers, the existing tools do not support such a mechanism. The most related work is about automated test input generation for Android, which is used to exploring useful information. The major approaches can be divided into three categories.

\textbf{Random Exploration Strategy} employs a random strategy to generate inputs for Android apps. Monkey ~\cite{Monkey} is the most frequently used tool based on the strategy which achieves a higher level of code coverage and error correction capabilities. However, it only generates UI event. Machiry \textit{et al.} presents Dynodroid ~\cite{Machiry2013Dynodroid}, which can trigger system events. Dynodroid selects the events that have been least frequently selected. Hu \textit{et al.}~\cite{Hu2011Automating} builds a system on top of Monkey for automatically detecting GUI bugs. Although randomly generating system events is easy to deploy, they would be highly inefficient, as there are too many events, and applications usually react to only few of them.

\textbf{Model-based Exploration Strategy} builds and uses a finite state machine of Graphical User Interface (GUI) model to generate events and systematically explores the behavior of the application. For example, Amalfitano \textit{et al.} presents GUIRipper~\cite{Amalfitano2012Using}, which later becomes MobiGUITAR~\cite{Amalfitano2014MobiGUITAR}. It uses DFS search strategy to dynamically construct a model of the app under test by crawling it from an initial state. Azim \textit{et al.}~\cite{Azim2013Targeted} presents a more abstract model in A3E-Depth-first. The model represents each activity as a single state, which results in missing some behaviors. Yang \textit{et al.} presents Orbit~\cite{Yang2013A}, which uses static analysis to extracts the set of events supported by the GUI of application, then dynamically explores the application by systematically executing these events. Choi \textit{et al.} designs SwiftHand~\cite{Choi2013Guided}, which optimizes the exploration strategy by minimizing restarts of the app while exploring to improve the speed. The Magic tool proposed by Nguyen \textit{et al.}~\cite{Nguyen2012Combining} is used to generate test cases for apps using a combination of model-based testing and combinatorial testing. Hao \textit{et al.} designs a novel framework PUMA~\cite{Hao2014PUMA}, which can be easily extended to implement any dynamic analysis on Android apps using the basic monkey exploration strategy. Liu \textit{et al.} designs DECAF~\cite{Liu2014DECAF}, incorporating a novel state equivalence prediction method to prioritize which paths to traverse in the UI graph for detecting structural fraud. The  prominent advantage of model-based strategy are low redundancy and the high code coverage, and it is well suited for reproducing task.

\textbf{Systematic Exploration Strategy} generates test cases for some special apps whose behavior can be revealed by providing specific inputs. For example, Azim \textit{et al.} presents A3E-Targeted~\cite{Azim2013Targeted}, which aims at maximizing the code coverage by different ways to build more systematic events. Anand \textit{et al.} presents ACTEve\cite{Anand2012Automated}, which symbolically tracks events. Merwe \textit{et al.} presents JPF-Android~\cite{Merwe2014Execution}, which explores all paths in an Android app, thus it can identify deadlocks and runtime exceptions. Systematic Exploration Strategy can facilitate symbolic execution and evolutionary algorithms to guide the exploration towards previously uncovered code, however, the poor scalability is the main shortcoming of this strategy.

%% file: section/tool.tex
\begin{figure*}[t!]
\centering
  \includegraphics[width=0.73\textwidth]{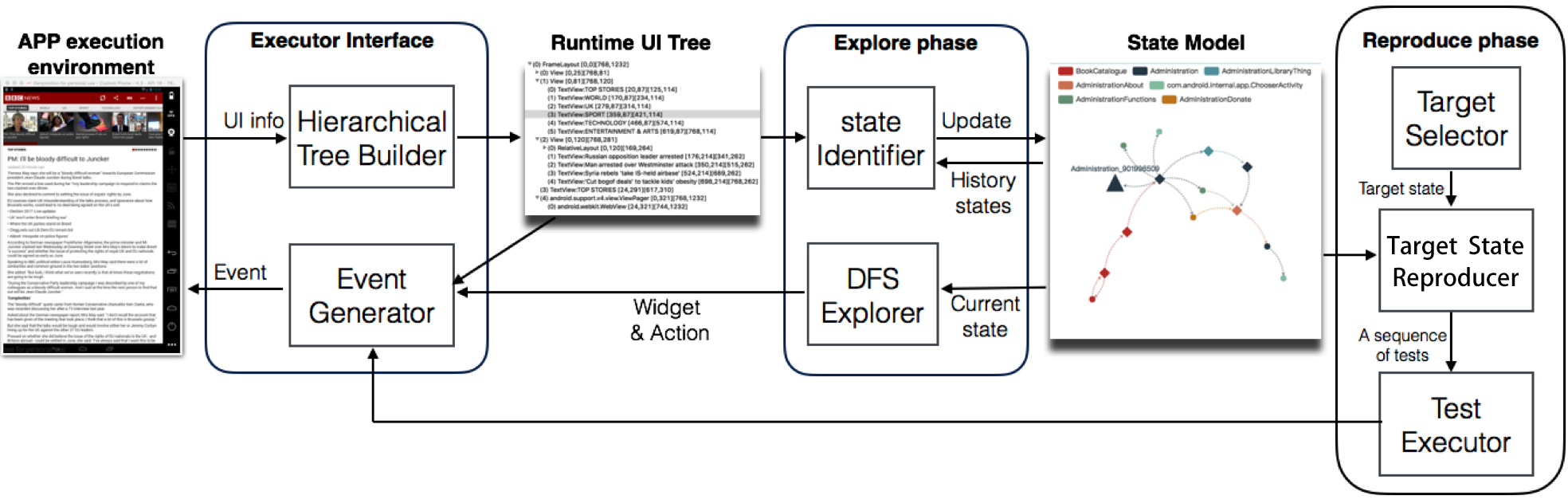}
  \caption{The overview of DroidWalker, which explores the app and generates a model. With the model, DroidWalker can generate reproducible test cases.}~\label{fig:workflow}
\end{figure*}

In this section, we describe the architecture of DroidWalker, as shown in Fig.~\ref{fig:workflow}. First, DroidWalker uses a model-based approach to explore as many app states as possible, while constructing a state model. We use UI structure to identify a state in a proper granularity. We also design an executor interface to generate flexible and dynamic-adaptive events. Second, developers can select a target state from the model so that test cases are generated to reproduce the state. DroidWalker traverses the model in a breadth first order to enumerate every possible event sequences to reach the target state, and executes these event sequences one by one. The rest of this section describes details of the design and implementation of each component.

\subsection{Executor Interface}

The executor interface connects with the app execution environment. It provides necessary functions for the higher level program modules. In our implementation, we build the executor interface on top of the robotium framework~\cite{robotium}, which is an Android test automation framework that has full support for android apps. The interface supports two major functions. On the one hand, such as extract structural UI information and design a data structure to represent hierarchical tree. On the other hand, the interface will be notified which widget should be executed by a certain action. Then the interface will locate the widget in the runtime and send the event.

\subsection{State Identifier}
DroidWalker models the GUI behavior of an Android app as a finite-state machine. In order to find a state identification criteria neither too fine-grained nor too coarse-grained, we propose to use the UI structure to identify states. This design is inspired by an empirical fact that regardless of the different assigned parameters, the UIs rendered by the same app behavior always share the same structure, and different app behaviors are usually different in the structure. For example, the pages of different restaurant details have the same structure, but the detail page and list page have obvious differences in the UI structure. In this way, we bring in some tolerance to the UI difference.

\begin{algorithm}
\footnotesize
    \caption{Computing structure hash of view tree.}
    \label{algo:structhash}
    \SetAlgoLined
    \KwIn{View $r$}
    \KwOut{Structure Hash $h$}
        \textbf{function}~TreeHash($r$)\\
        $str \leftarrow r.viewTag$\\
        \If{$r.InstanceOfWebview()$} {
                $r.setChildren \leftarrow r.parseHTML()$\\
        }
        \If{$r.hashChildren()$} {
            $children \leftarrow r.getChildren()$\\
            \ForEach{$c \in children$} {
                $c.hash \leftarrow TreeHash(c)$\\
            }
            $children \leftarrow SortByHash(r.getChildren())$\\
            \If{$r.InstanceOfListview()$} {
                $children \leftarrow children.unique()$\\
            }
            \ForEach{$c \in children$} {
                $str \leftarrow str + c.hash$
            }
        }
    \Return{$hash(str)$}
\end{algorithm}

In Android, all the UI widgets are organized in a structure of hierarchy tree. To rapidly compare two UI's structural difference, we design a recursive bottom-up algorithm~\ref{algo:structhash} to encode the structure into a hash value, and use the hash to distinguish different UI's structure. The algorithm is recursive with a view $r$ as input. If $r$ does not have children, the result is only the string hash of $r$'s view tag (Line 2). If $r$ has children (Line 6), then we use the algorithm to calculate all the hash values of its children recursively (Lines 8-10). Then, we sort $r$'s children based on their hash values to ensure the consistency of the structure hash, because a view's children do not keep the same order every time(Line 11). In many cases, there exist a list-view or recycle-view that contain multiple views with the same structure. Under such circumstances, we should count the items with the same structure hash by only once (Line 13). Next, we add each children's hash together with the view tag, forming a new string (Line 16), and finally return the string hash (Line 19). Given the root view of the tree, the algorithm returns a structure hash of the view tree. The hash can be used as an identifier of a UI state.

This recursive structural encoding method can be extended to Web elements. Since increasing number of embedded WebViews appear in the Android apps, we also incorporate the Web elements inside WebView component into the hierarchical tree (Line 4) rather than treat the WebView as a leaf node.

\subsection{DFS Explorer}

Similar to the previous model-based app-exploration tools, we use a depth first strategy to explore the app. For each state, we extract all the potential widgets that have event listeners, and then systematically generate the event of each widget. Next, we check whether the event brings the app to a new state, by comparing its structural hash with all the other states in the state model. If a new state is identified, we recursively apply the exploring algorithm on the new state. When the exploration on this state terminates, we execute a backtrack method to the previous state.

\begin{algorithm}
\footnotesize
    \caption{DFS Explore}
    \label{algo:explore}
    \SetAlgoLined
    \KwIn{Current UI Tree $tree$, State Model $model$}
        \textbf{function}~DFS\_Explore($A$, $model$)\\
        \ForEach{$widget \in tree.getClickableWidget()$} {
            $GenerateEvent(widget)$\\
            $currentTree \leftarrow BuildTree()$\\
            \If{$it is a UI state not yet explored$} {
                $model.add(currentView)$\\
                $DFS\_Explore(currentTree, model)$\\
                $backTrack(tree, currentTree)$\\
            }
        }
\end{algorithm}

Although depth-first search has been widely adopted by various app-exploration tools, there is an efficiency problem related to backtracking, i.e. returning to the previous state and continuing to execute the remaining events of that state. The challenge is because the \emph{goBack} method provided by Android system only goes back to the previous activity rather than UI state. Therefore, a common strategy used by existing tools is to restart the app, and re-send the events from the initial state. This strategy can somehow achieve backtracking, but it faces two major drawbacks. On one hand, due to the dynamic nature of Android apps and the difficulty of accurately replaying, sometimes the reinstall method may not successfully return to the correct state. On the other hand, frequently restarting the app may cost dramatic amount of time, which compromises the exploration efficiency. Especially in the circumstance when we use the fine-grained UI states, a replay trace may include hundreds of states, and considering a precise UI replay always need to wait for some time for the UI state to stabilize.

To achieve the precise backtracking with minimal time cost, we propose to take the advantage of Android's intent mechanism. We record the intent information when transiting between activities at runtime. Whenever the executor backtracks to a previous state, it sends the intent of the target state's activity, and check if the current state is in the stack. If so, the executor will resend the UI traces in order to get the target state.

\begin{algorithm}
\footnotesize
    \caption{backTrack}
    \label{algo:back}
    \SetAlgoLined
    \KwIn{Target State $T$, Current State $S$}
        \textbf{function}~backTrack($T$, $S$)\\
        $stack \leftarrow T.UIStack$\\
        \If{$stack.activity \ne getCurrentActivity()$} {
            $SendIntent(stack.intent)$\\
        }
        $S \leftarrow getCurrentState$\\
        \ForEach{$state \in stack.states$} {
            \If{$Similarity(S, state) > threshold$} {
                $stack.ReplayTracesFrom(state)$\\
            }
        }
\end{algorithm}

\subsection{Target State Reproducer}
After the exploration, a state model is constructed. Developers can select any state to be reproduced. DroidWalker supports user customized detection function, such as malevolence, exception, or faults, by which significant states can be provided for developers.

Since the model also records the necessary event information, such as which widget should be executed and which action should be sent, the event can be regenerated between arbitrary two states. Therefore, the key idea of generating reproducible test cases is to find the potential paths from the entry state of the model to the target state. This can be done by a standard bread-first search. Started from the target state, the algorithm enumerates every potential state path, and sorts them by their length. These paths are a sequence of generated test cases.

\subsection{Test Executor}
After a sequence of tests are generated, the executor will launch these test cases one by one, by extracting the states and generate the events to each states. The executor will check at each step whether the correct state is reached, ensuring the authenticity of the test cases. Although the assigned hash value can distinguish nodes with different structure, its ability to tolerate app change is quite low. In the reproducing phase, we need a more adaptive approach to identify current state.

Due to the dynamic nature of android app, the UI may sometimes change even when the same sequence of events are executed, and a trivial UI change may result in a totally different structural hash value. For example, when a test case to explore news page is generated, a notification about the news update may appear on top of the screen. Since it is only a trivial UI change and does not influence the majority of the other functions, we should tolerate this difference and continue executing. Therefore, in order to make our approach more scalable, it is necessary to incorporate a structural similarity criteria. Only when a UI's similarity difference to the previous state is below the pre-defined threshold should we assign a new model state. The algorithm~\ref{algo:diff} accept two arbitrary view node and a threshold as inputs.

\begin{algorithm}
\footnotesize
    \caption{Structual Similarity between two hierarchical UI trees.}
    \label{algo:diff}
    \SetAlgoLined
    \KwIn{View $s$, View $t$, threshold $thd$}
    \KwOut{Structual Similarity $sim$}
        \textbf{function}~Similarity($s$, $t$)\\
        \If{$s.hash = t.hash$} {
            \Return{$1$}\\
        }
        \If{$s.tag \neq t.tag$} {
            \Return{$0$}\\
        }
        $hits \leftarrow 1$\\
        \ForEach{$sc \in s.getChildren()$} {
            \ForEach{$tc \in t.getChildren()$} {
                $tmp \leftarrow Difference($sc$, $tc$)$\\
                \If{$tmp > threshold$} {
                    $hits \leftarrow tmp * tc.count$\\
                    $break$\\
                }
            }
        }
        \Return{$2 * hits / (s.count + c.count)$}
\end{algorithm}

If the two UI nodes share the same hash value, their structure are the same, so the similarity equals 1 (Line 2). Otherwise, there must be some difference between them, so we enumerate the children of these two nodes, and calculate the similarity of these children (Line 9-17). To reduce the complexity, we will stop traversing if a children pair reach the threshold (Line 12). The similarity is the shared nodes divided by the total nodes (Line 18).

In this way, the state our tool identified is just appropriate for reproducing test case.

%% file: section/Usage_scenarios.tex
\begin{figure}[t!]
\centering
  \includegraphics[width=0.45\textwidth]{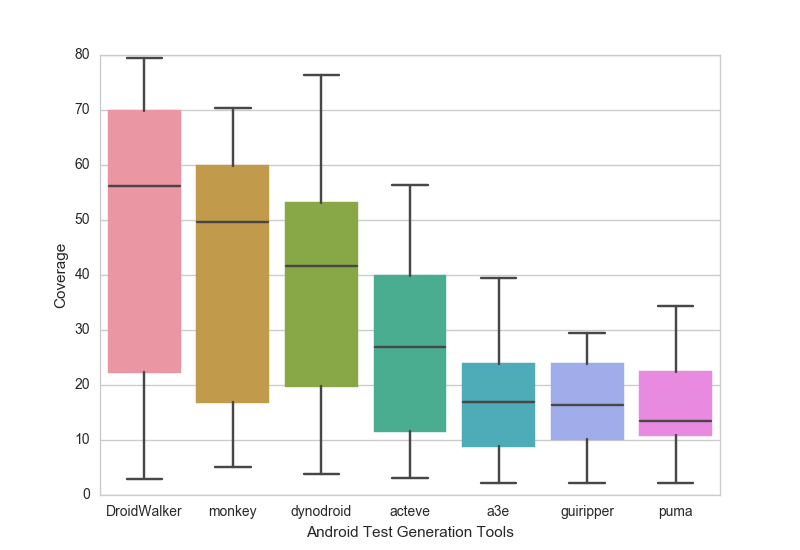}
  \caption{Boxplot of the line coverage result across 20 benchmark apps.}~\label{fig:res}
\end{figure}

\begin{table*}[t!]
\scriptsize
\caption{Activity and method coverage result across five commercial app experiment, comparing with Monkey.}
\label{tab:seven_table}
\centering
\begin{tabular}[t]{|c|c|c|c|c|c|c|}
    \hline
    \textbf{App} & \textbf{Category} & \textbf{\tabincell{c}{Total\\ Activities}} & \textbf{\tabincell{c}{Activity Coverage\\ of DroidWalker}} & \textbf{\tabincell{c}{Activity Coverage\\ of Monkey}} & \textbf{\tabincell{c}{Method Coverage\\ of DroidWalker}} & \textbf{\tabincell{c}{Method Coverage\\ of Monkey}} \\ \hline
    \textbf{accuweather} & weather  &    37    &    11    &   8   &   2314     &     1786 \\
    \hline
    \textbf{lionbattery} & tools    &    100   &    29    &   23  &   2034     &     1317 \\
    \hline
    \textbf{iconology} & comics     &    41    &    22    &   18  &   3039     &     2253 \\
    \hline
    \textbf{fibit} & Health         &    92    &    28    &   21  &   6702     &     4927 \\
    \hline
    \textbf{vyapar} & Business      &    88    &    39    &   36   &  2543     &     1759\\
    \hline
\end{tabular}
\end{table*}

DroidWalker is designed to automatically generate reproducible test cases for Android apps. The current implementation offers a command-line tool and a visualized webpage. This section demonstrates three main usage scenarios of DroidWalker: (1) automate app exploration, (2) customize detection functions (3) reproduce target state.

\subsection{Automate App Exploration}

The first usage scenario is using DroidWalker to explore app under test automatically. Users should follow steps below:
\begin{enumerate}
\item Prepare the apk of the app to be explored. Then run a python script to repackage and build the app. The script can be configured with command-line arguments below, the repackaged project is located in the <lists> folder under the working directory:

\begin{tiny}
\begin{lstlisting}[ language=C]
python buildAndResigned.py APK\_Address
[--ip<address>] [--port<port>] [--output<graph|report|both>]
\end{lstlisting}
\end{tiny}

* [--ip<address> --port<port>]: optional, specify ip and port of the server used to upload exploring output, default to be localhost:5000
* [--output<graph|report|both>]:  optional, specify whether to output coverage report and the content of output, including a graph which is a visualization of the constructed GUI model, a progressive coverage plot over time. The default argument is both.

The output information will be visualized on the webpage. If the source code of the app under test is available, users can choose to instrument the app with Emma to get a more detailed coverage report or the coverage information will be generated using logcat and tracedump offered by Android SDK.

\item Start the server of visualized webpage using the command line: \emph{python server.py}, the console will output the specific listening port of the server.

\item Open Android emulator or connect with Android devices, then change directory to <lists/TestAPP/app/build/outputs/apk>. Start exploring the app using the command line: \emph{./runTest.sh}

\item Open a browser and enter the listening address of the server, and start monitoring the exploring output.
The webpage consists of a tab panel. By clicking the tab, user can switch between graph and coverage report. The webpage also demonstrates snapshots of the app, each of which corresponds to a specific node in the graph.

\end{enumerate}

To assess the efficiency of DroidWalker when applied to automate app exploration, we conduct two experiments to address the following two questions: first, whether our tool is superior than other automated exploring tools on a standard benchmark. Second, whether the tool can be applied to highly complex commercial apps.

For the first question, in order to be comparable, we use a set of mobile app benchmarks collected in the work done by Choudhary \textit{et al.}~\cite{Choudhary2016Automated}, and concerned both the code coverage and the coverage time as criteria to evaluate our tool.

The experiment setup is as follows:

1) Experiment 1: We ran our tool in the same setting as Choudhary \textit{et al.}~\cite{Choudhary2016Automated}, using the Ubuntu virtual machine configured with 2 cores and 6GB RAM, and an android emulator was configured with 4GB RAM. Each tool was allowed to run for 1 hour on each benchmark app. We chose Android SDK version 18 (Jelly Bean) to evaluate our tool based on the compatibility of most benchmark apps. To avoid unexpected side effects between tools and app, the emulator is destroyed after finishing each run. We repeated exploring each app for 3 times. From Fig.~\ref{fig:res}, we can see that line coverage of our tool is higher than those of the state-of-art tools, especially the model-based tools, such as A3E,GUIRipper and PUMA.

Although exploring a common set of apps is necessary for comparison, the fact that most apps in benchmarks are small and has simple logics indicates the result could not represent the performance of exploring tool on modern commercial apps. Thus, we select ten top apps from different categories on the Google Play and use the same two criteria to evaluate those apps. The experiment setup is as follows:

2) Experiment 2: Five top apps in Google Play are chosen to be tested. Because the applicaitons are more complicated, we chose Genymotion, an optimized android emulator to perform the experiment. The emulator settings are identical to Experiment 1. From Table~\ref{tab:seven_table}, we can see that our tool outperforms monkey even on complex commercial apps.

\subsection{Customize Detection Function}

The second usage scenario is using DroidWalker to customize detection function which can be called during exploration. User can write their own class under <config> directory. The class should be inherited from \texttt{Customize} class and override three methods--input, output and operate. Add the following argument while repackaging and building the app:
$*[--configure-class <Customized\_class>]$

The Customize class contains an outputContent field. The input method is designed to accept existing data in string format provided by users. The operate method is designed to be the central customized detection function. The output method is designed to deal with output, which is default to be sent to the server of visualized webpage.

Customized class will be instantiated inside the testing program and called during exploration. User can customize their own class to detect advertisements, code defects or code vulnerability. To demostrate customizing testing function, we write an exception collection class as example. The configuration of Android emulator is the same as experiment 2. The class example is as follows:

\begin{tiny}
\begin{lstlisting}[language=java, label = {code:except}]
public class ExceptionCollector extends Customize {
  public void operate(){
    Process process = Runtime.getRuntime().exec("logcat -d PACKAGE_NAME:I");
    BufferedReader bufferedReader = new BufferedReader(
        new InputStreamReader(process.getInputStream()));
    StringBuilder log = new StringBuilder();
    String line = "";
    Boolean flag = false, has_except = false;
    while ((line = bufferedReader.readLine()) != null) {
      if (line.contains("Exception")){
        flag = has_except = true;
        log.append(line + "\n");
      }else if(flag){
        if (line.contains("at")) log.append(line + "\n");
        else flag = false;
      }
    }
    Runtime.getRuntime().exec("logcat -c");
    if (has_except) outputContent = log.toString();
  }
}
\end{lstlisting}
\end{tiny}

then repackage and build the project using:

\begin{tiny}
\begin{lstlisting}[language=python]
python buildAndResigned.py APK --configure-class ExceptionCollector
\end{lstlisting}
\end{tiny}

We applied our tool to Fitbit, a top health care appliaction on Google Play. The tool recorded two exceptions during exploration. The first one is java.lang.UnsupportedOperationException, which suggests the testing program attempted to open a session that has a pending request to the user's FaceBook account. The second one is java.io.FileNotFoundException, which was raised when the testing program attempted to take pictures or select photo from album for profile. The trace information is as follows:

\begin{tiny}
\begin{lstlisting}[language=python]
failed to get https://android-api.fitbit.com/1/user/5NZZKF/badges.json
from cache  java.io.FileNotFoundException:
/data/data/com.fitbit.FitbitMobile/cache/datacache/https%3A%2F%2Fandroid
-api.fitbit.com%2F1%2Fuser%2F5NZZKF%2Fbadges.json:
 open failed: ENOENT (No such file or directory)
    at libcore.io.IoBridge.open(IoBridge.java:409)
    at java.io.FileInputStream.<init>(FileInputStream.java:78)
    at com.fitbit.data.repo.q.a(SourceFile:71)
    at com.fitbit.serverinteraction.ServerGateway.b(SourceFile:811)
    at com.fitbit.serverinteraction.ServerGateway.a(SourceFile:780)
    at com.fitbit.serverinteraction.PublicAPI.a(SourceFile:352)
    at com.fitbit.serverinteraction.PublicAPI.b(SourceFile:2376)
    at com.fitbit.data.bl.BadgesBusinessLogic.a(SourceFile:171)
    at com.fitbit.profile.ui.badges.a.c(SourceFile:48)
    at com.fitbit.profile.ui.badges.a.b_(SourceFile:18)
    at com.fitbit.util.bd.loadInBackground(SourceFile:36)
     ...
\end{lstlisting}
\end{tiny}

\subsection{Reproduce Target State}
The third usage is using DroidWalker to reproduce target states. DroidWalker will number each UI status. User can relate UI number with its corresponding snapshot and customize UI number sequence. The app uses Breath-First-Search strategy to generate a sequence of test cases, which are sorted by their event length, and these test cases will be executed one by one.

Users can detect severe code errors when DroidWalker encounters a crash. After revising related code, user can enter the UI number causing the crash to check whether the error has been fixed. Users can also make use of the customized detection functions, repeating test cases and checking the detection results to comfirm app behaviors.

\begin{figure}
\begin{minipage}[t]{0.47\linewidth}
\centering
\includegraphics[width=1.55in]{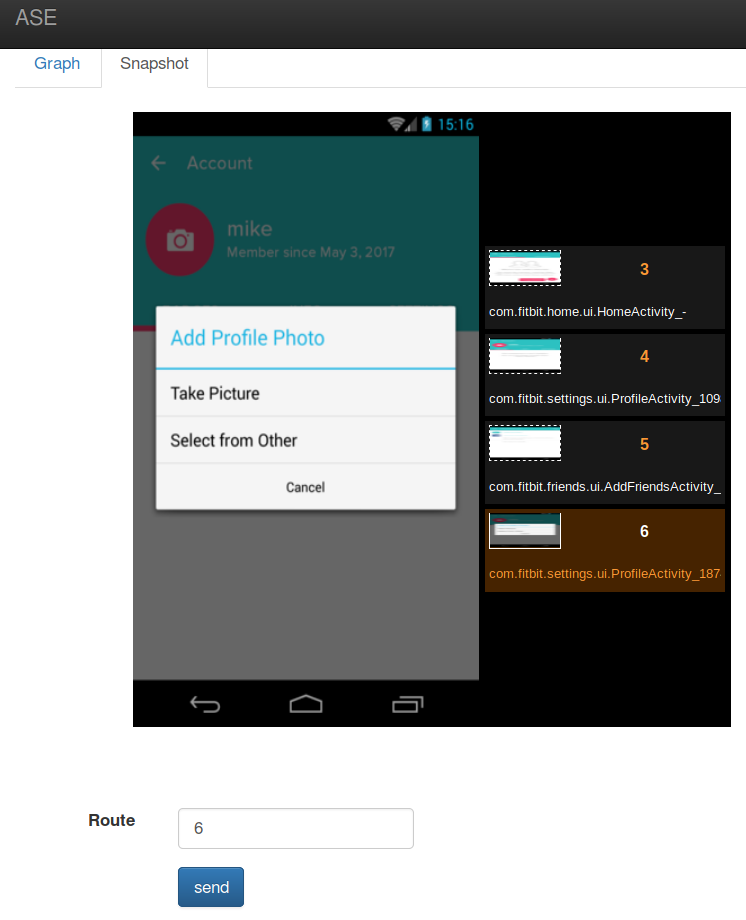}
\caption{Target state selector.}
\label{fig:side:a}
\end{minipage}%
\begin{minipage}[t]{0.47\linewidth}
\centering
\includegraphics[width=1.8in]{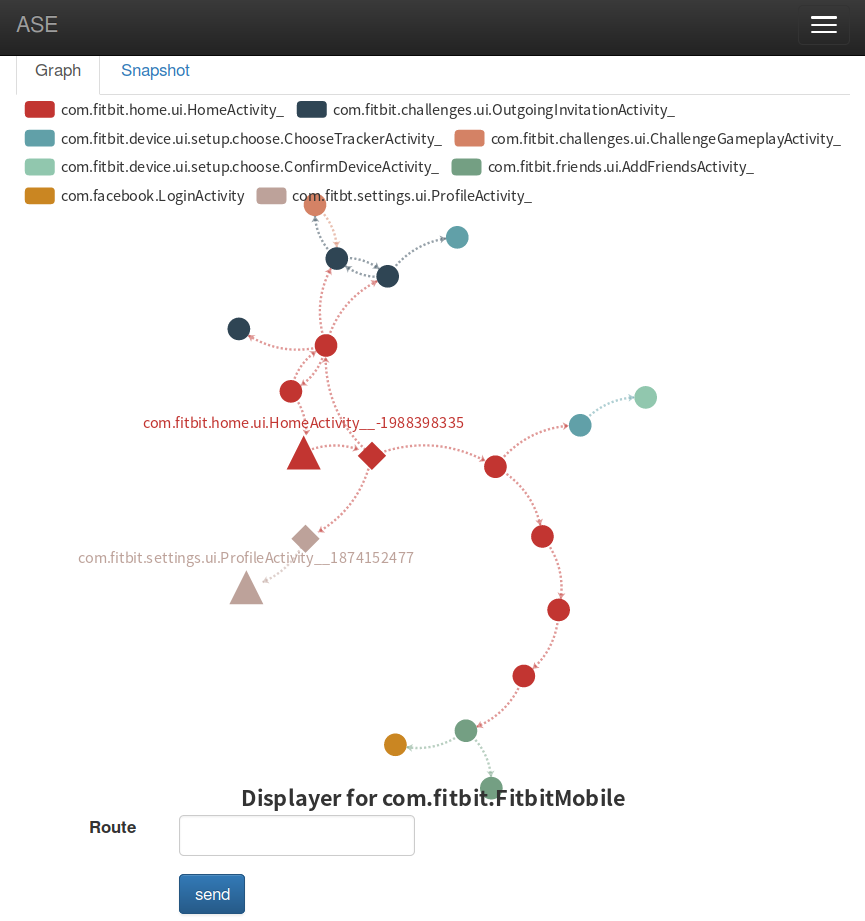}
\caption{State model displayer.}
\label{fig:side:b}
\end{minipage}
\label{fig:side}
\end{figure}

For demonstratation, We use the tool to reproduce the java.io.FileNotFoundException exception recorded by the detector. On the webpage, We can find out the corresponding snapshot number is 6, as is shown in Fig.~\ref{fig:side:a}. Enter 6 in the input field below the graph and click \emph{send}.

The testing program successfully search the shortest path to the profile page, as is shown in Fig.~\ref{fig:side:b}, and start exploring again. Once the testing program click on \emph{Take Picture} or \emph{Select from other}, the exception repeated again.

%% file: section/conclusion.tex
This paper presents DroidWalker, an automatic exploration tool that can generate reproducible test cases for Android apps. DroidWalker can efficiently explore the app and build a dynamic-adaptive model. Developers can configure customized monitoring tasks to examine the app behaviors. After exploration, DroidWalker can generate reproducible test cases to all the states in the model, thus facilitating many real-world test tasks such as failures reproduction and regression test.